# Nanomechanical Measurements of Magnetostriction and Magnetic Anisotropy in (Ga,Mn)As

S. C. Masmanidis[1], H. X. Tang[1], E. B. Myers[1], Mo Li[1],
K. De Greve[1,2], G. Vermeulen[1,2], W. Van Roy[2] and M. L. Roukes[1]

[1]*Condensed Matter Physics 114-36, California Institute of Technology, Pasadena, California 91125*
[2]*IMEC, Kapeldreef 75, B-3001 Leuven, Belgium*

A (Ga,Mn)As nanoelectromechanical resonator is used to obtain the first direct measurement of magnetostriction in a dilute magnetic semiconductor. Field-dependent magnetoelastic stress induces shifts in resonance frequency that can be discerned with a high resolution electromechanical transduction scheme. By monitoring the field dependence, the magnetostriction and anisotropy field constants can be simultaneously mapped over a wide range of temperatures. These results, when compared with theoretical predictions, appear to provide insight into a unique form of magnetoelastic behavior mediated by holes.

The dilute magnetic semiconductor (DMS) (Ga,Mn)As has been extensively studied for its promising spintronics applications [1,2]. Among its properties is the dominant role of growth-induced strain upon the material's magnetic alignment. A change from compressive to tensile strain is known to flip the moments from in-plane to out-of-plane [1,3]. This is consistent with the inverse magnetoelastic effect, but the magnetoelastic coupling parameters that quantify this behavior have remained elusive. Here, instead of relying on strain from the growth, we demonstrate a scheme to observe magnetostriction directly in a resonant nanoelectromechanical system (NEMS). The piezoresistive, piezoelectric and elastic properties of (Ga,Mn)As should be similar to those of doped GaAs, enabling straightforward electromechanical actuation and transduction [4,5]. Measurable frequency variations occur as a magnetoelastic doubly-clamped beam resonator is stretched or compressed in an applied magnetic field [6,7]. Furthermore, the angular dependence of the observed magnetostriction enables measurement of magnetic anisotropy.

The material is grown epitaxially on an (001) GaAs substrate, beginning with a 1 μm $Al_{0.8}Ga_{0.2}As$ sacrificial layer, followed by 50 nm high temperature and 50 nm low temperature GaAs, and finally 80 nm unannealed $Ga_{0.948}Mn_{0.052}As$ with a Curie temperature of ~57 K. The spontaneous magnetization is expected to lie in the growth plane due to a compressive strain from the substrate and demagnetization effects. Electron beam lithography is used to define the device profile, which is subsequently covered with a titanium etch mask. Next, argon ion milling removes all magnetic material not protected by the mask. A 30 nm-thick gold side gate is deposited 0.7 μm away from the beam after another lithography step. Finally, a rectangular resist window is patterned to expose the sacrificial layer, which is selectively removed along with the remaining titanium mask in dilute hydrofluoric acid. The resulting suspended structure, shown in the inset of Fig. 1(a), has dimensions of (L, w, t)=(6, 0.5, 0.18) μm, with its longitudinal axis oriented along the [110] crystallographic direction. The sample is mounted in a liquid helium cryostat in vacuum, and surrounded by a 3-axis, 10 kOe superconducting magnet. The two-terminal device resistance at 4.2 K is 80 kΩ. By applying an AC voltage on the side gate, a piezoelectric dipole interaction [4] drives the beam out-of-plane in its fundamental resonance mode.

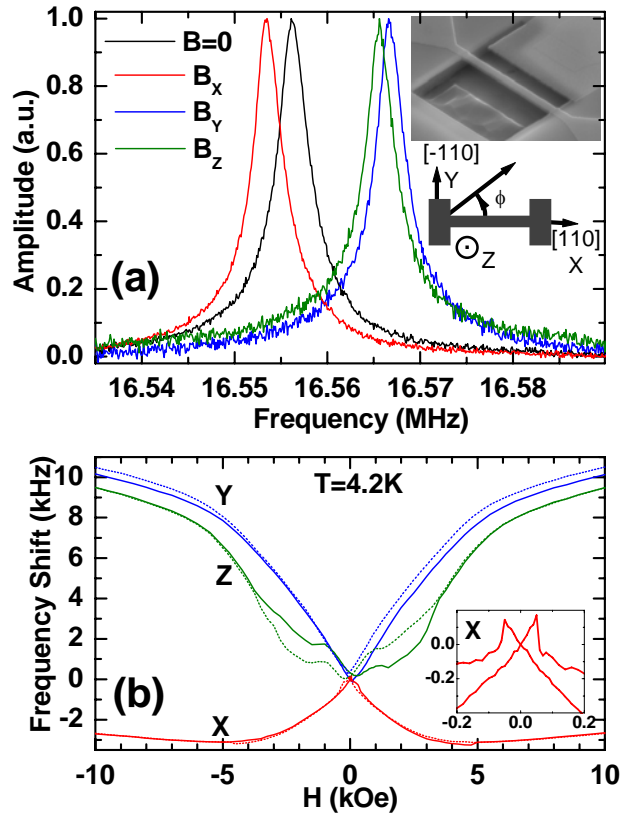

FIG. 1 (color). (a) Frequency response of the device for various 10 kOe field directions (Z is out of the plane). Amplitude is normalized for clarity. INSET. SEM image of the suspended (Ga,Mn)As beam and Au side gate with the axis directions. (b) Frequency shift dependence on field magnitude and direction. Dotted curves are for reversed field sweep direction. INSET. Expanded view of shifts in a field along X, showing domain wall-induced transitions.

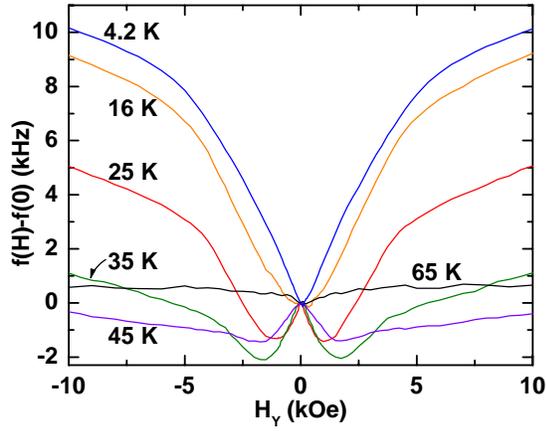

FIG. 2 (color). Frequency shift dependence on a field along $\phi_H = 90°$ for different temperatures.

Transduction is carried out via a sensitive piezoresistive down-mixing scheme [5] with the signal detected using a lock-in amplifier, following low noise preamplification.

The piezoresistive frequency response along different magnetic field directions is shown in Fig. 1(a). At 4.2 K and zero external field, the device resonates at 16.56 MHz with a quality factor of 6,300. The resonance is tuned up or down, depending on whether the applied field is aligned parallel or orthogonal to the [110] direction. Note that torque-induced frequency shifts [8] that are prevalent in larger devices, are expected to have a negligible impact on the resonance frequency of the smaller, stiffer DMS NEMS device studied here.

The device can be operated as part of a phase-locked loop (PLL), enabling real-time resonance frequency tracking with a resolution of 4 parts in $10^7$ at 4.2 K. Figure 1(b) shows the frequency shifts as a function of field along the three principal beam directions. Experimentally one can identify three distinct tuning regions corresponding to low (<100 Oe), intermediate (100-5000 Oe) and high (>5000 Oe) field behavior. In the low field region magnetization reversal appears to proceed via domain wall displacement, characterized by abrupt changes in frequency as shown in the Fig. 1(b) inset. The intermediate region apparently coincides with coherent moment rotation described by the Stoner-Wohlfarth model [9]. Note that hysteresis can be found in the first two regions. The magnetization reversal is complete beyond ~5000 Oe, but frequency continues to increase with field. This last observation is consistent with a forced magnetostriction effect [10]. The field dependence measurements are extended to higher temperatures in Fig. 2. For clarity, only the results for a field oriented in-plane and orthogonal to the beam, are displayed. The magnitude of the frequency shift decreases with temperature, and a small paramagnetostrictive effect persists above the Curie point up to at least 65 K. The slope of the high field, linear part of the curves provides a rough gauge of the forced magnetostriction. We find that this effect decreases with temperature and vanishes at around 60 K. The concurrence of the onset of forced magnetostriction with ferromagnetic ordering in our sample confirms that this phenomenon is intrinsic to (Ga,Mn)As, however, we currently lack a clear understanding of this effect in DMS. It is also noteworthy that starting at 20 K a downshift in frequency occurs at low and intermediate fields. This behavior becomes increasingly pronounced at higher temperature, before disappearing altogether in the paramagnetic regime.

To gain a better understanding of magnetoelastic coupling we use a PLL to track resonance frequency shifts in a constant field that is rotated in the plane of the device. The results are presented in polar form in Fig. 3, with the field chosen as 5 kOe such that magnetization reversal occurs purely by rotation, *i.e.* no domain wall displacement. Forced magnetostriction effects will also be curtailed at this intermediate field value. The field angle $\phi_H$ is measured with respect to the [110] direction. Between 4.2 K and 20 K we see a two-fold symmetry in the angular dependence of the resonance frequency, which is maximized along 90° and 270°, and minimized along 0° and 180°. However, by 25 K, additional symmetry emerges in the form of two new peaks at 0° and 180°. The new peaks grow with temperature relative to the original pair, and above 35 K they are the dominant feature of angular dependence in Fig. 3. The onset of this behavior resembles that of the low field frequency shifts seen in Fig. 2, and suggests qualitative changes in magnetoelastic coupling and anisotropy occur as a function of temperature.

We now attempt to extract quantitative information from the data in Fig. 3. The angular dependence of the frequency shifts is modeled after the magnetostriction equation [10,11] containing the first order terms $\lambda_{100}$ and $\lambda_{111}$ corresponding to volume-conserving deformations, and a second order term $h_3$ corresponding to a volume-changing deformation. The two-fold or four-fold symmetry found in the polar plots can be uniquely and unambiguously described by first and second order magnetostriction. Combining this model with the stress-strain relation gives the following expression for the excess longitudinal magnetoelastic stress on the beam relative to the zero field state:

$$\sigma_{ML} = (\lambda_{100}/4)(c_{11}-c_{12}) + (3\lambda_{111}/4)(c_{11}-c_{12})\cos 2\phi_M \\ + (h_3/4)(c_{11}+2c_{12})\cos^2 2\phi_M \quad (1)$$

where $\phi_M$ is the in-plane magnetization angle. The inclusion of the second order term is necessary to explain the four-fold symmetry seen above 20 K. The precise elastic constants of $Ga_{0.948}Mn_{0.052}As$ are unknown but assumed to be very similar to those of GaAs: $c_{11}$=121.6 GPa, $c_{12}$=54.5 GPa and Young's modulus E=86 GPa [12]. The zeroth order correction factor to frequency from a longitudinal stress is given by $(1+\sigma_{ML}L^2/4Et^2)^{1/2}$ [13]. In practice we rely on a finite element simulation to model the beam's dynamics upon stressing the 80 nm-thick magnetic top layer of the beam. For the range of relevant magnetoelastic strain, the model predicts a linear stress-frequency gauge factor that for the specific device geometry used here is equal to $\Delta f / \Delta \sigma_{ML}$=-13.9 Hz/kPa.

We find it necessary to include a cubic and in-plane uniaxial magnetic anisotropy in our model in order to quantitatively describe the curvature of the angular dependence of frequency in Fig. 3. The Stoner-Wohlfarth model is used to couple the magnetoelastic stress equation, expressed in terms of the magnetization angle, to the data which is measured with respect to the field angle. The corresponding minimum free energy condition is [14]:

$$H_{KU} \sin 2\phi_M - (H_{K1}/2)\sin 4\phi_M + 2H \sin(\phi_M - \phi_H) = 0 \quad (2)$$

The first order in-plane uniaxial and cubic anisotropy fields are given by $H_{KU}=2K_U/M$ and $H_{K1}=2K_1/M$. We ignore the magnetoelastic second order anisotropy contribution to free energy, because this varies as $\sim E\lambda^2 \cos^4 \phi_M$ and can be disregarded for small $\lambda$. With this assumption, Eqns. 1 and 2 are effectively coupled via a single variable $\phi_M$, enabling extraction of the magnetostriction and anisotropy parameters by a straightforward best fit analysis. First, the three magnetostriction constants are obtained by applying the stress-frequency gauge factor on Eqn. 1 and fitting to the frequency shift maxima and minima that occur in Fig. 3 at multiples of $\phi_M$=90°. The magnetization is independent of anisotropy along these field directions. To obtain $H_{KU}$ and $H_{K1}$, Eqn. 2 is fed into Eqn. 1 after being solved with trial anisotropy constants, and the procedure is iterated to produce the best fit to the data. We find this method quantitatively explains the angular dependence of the frequency shifts over the entire ferromagnetic regime, up to the Curie temperature of ~57 K. The experimentally derived magnetostriction and magnetic anisotropy field parameters are plotted in Figs. 4(a,b). At 4.2 K, the first order magnetostriction constants of $Ga_{0.948}Mn_{0.052}As$ along [100] and [111] are $\lambda_{100}$=-11.3 ppm and $\lambda_{111}$=8.1 ppm, respectively. The corresponding magnetoelastic strain coefficient along the [110] direction is given by [10] $\lambda_{110}=\lambda_{100}/4+3\lambda_{111}/4$=3.2 ppm. The measured anisotropy fields are the same order of magnitude as those from studies on bulk GaMnAs films [15,16].

The temperature dependence displayed in Fig. 4 reveals an intricate coupling between magnetostriction and magnetic anisotropy. At 4.2 K both $H_{KU}$ and $H_{K1}$ are positive and $H_{K1}>H_{KU}$, indicating the magnetization lies in the (001) plane with cubic easy axis symmetry close to [100] and [010]. The presence of $H_{KU}$ will tilt the moments in the direction of [110] by $0.5\sin^{-1}(H_{KU}/H_{K1})$=9° [14]. The magnitude of the tilting angle gradually increases and by 25 K, complete realignment along [110] has occurred accompanied by a changeover to uniaxial easy axis symmetry. It is notable that the same transition is observed elsewhere [15]. The span between 30 and 40 K is marked by significant qualitative changes in all measured parameters. Specifically, $\lambda_{100}$ and $\lambda_{111}$ change sign while $H_{KU}$ and $h_3$ attain their respective local maximum and minimum values. The trend of $H_{K1}$ suggests it changes sign at around 40 K. In spite of this, the device retains its uniaxial easy axis symmetry along [110], since $|H_{KU}|>|H_{K1}|$.

The cubic anisotropy in DMS is predicted to change sign with variations in hole density, owing to the central role of carriers in mediating ferromagnetism [17,18]. In Fig. 4b, we see this occurring to $H_{K1}$ with increasing temperature. An apparently related phenomenon is the sign change in cubic [19] and uniaxial anisotropy [20] observed upon raising temperature. These observations may be the result of a thermally-driven increase in hole density. Like anisotropy, magnetostriction arises from interactions between neighboring magnetic moments, and as a consequence, magnetostriction is also expected to be coupled to the carrier density. The parameters $\lambda_{100}$ and $\lambda_{111}$

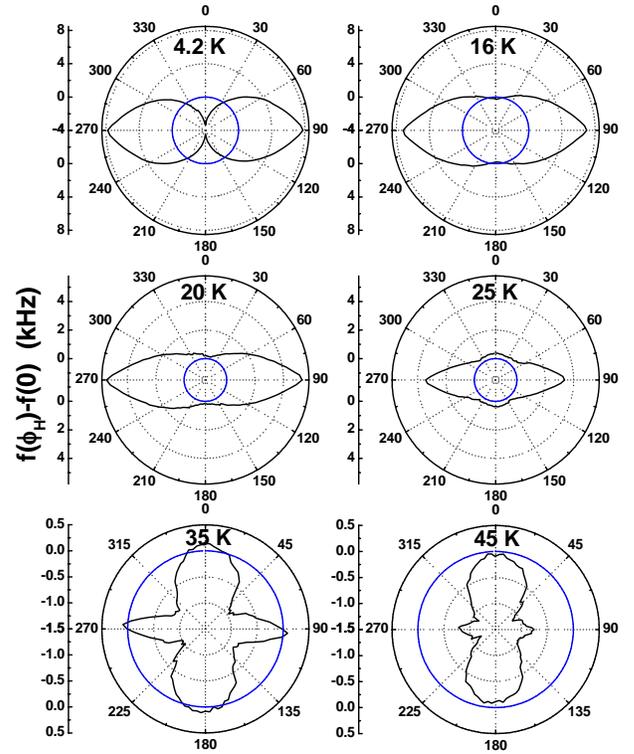

FIG. 3 (color). Polar plot of angular dependence of frequency shifts for an in-plane 5 kOe field. The field angle is measured relative to [110]. Note the change of scale on the axes. The $\Delta f$=0 reference line is shown in blue.

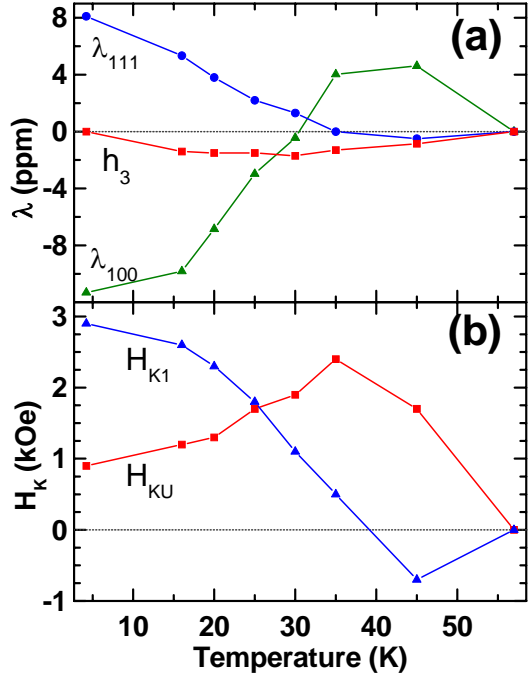

FIG. 4 (color). (a) First and second order magnetostriction constants of $Ga_{0.948}Mn_{0.052}As$. (b) First order cubic and in-plane uniaxial anisotropy fields.

both change sign and follow a trend with increasing temperature resembling that of $H_{K1}$; this further supports the above hypothesis. The dependence of valence band energy on strain leads us to expect a sensitive interplay between hole density, magnetic anisotropy, magnetostriction and other sources of strain. In particular, the thermal expansion of undoped GaAs is negative and locally minimized in the region of 32 K [21] and could play a role in mediating the behavior seen in Fig. 4. Moreover, stresses may be amplified in suspended nanostructures such as the one used here because of additional free surfaces and an increased surface to volume ratio.

Inverse magnetostriction is expected to impact magnetic anisotropy. Its contribution can be approximated as a uniaxial anisotropy field of the form $H_{KU}^\sigma \sim -3\lambda_{ii}\sigma_{ii}/M$ [10,17] where $\sigma_{ii}$ is the net applied stress on the beam along a specified direction. A tetragonal distortion of $Ga_{0.948}Mn_{0.052}As$ grown on GaAs would lead us to expect a cubic in-plane anisotropy along the [110] and equivalent axes, but it has been argued from studies of ultrathin Fe grown on InAs [22] that surface reconstruction leads to asymmetric strain relaxation favoring uniaxial anisotropy. Finally, demagnetizing fields are also expected to play a role [23], but decrease with temperature and thus cannot account for the observed rising trend below 35 K.

In summary, we have employed a nanomechanical resonator to measure the magnetostriction constants and magnetic anisotropy fields of (Ga,Mn)As. We observe a gradual temperature-driven crossover at 25 K from in-plane cubic to uniaxial easy axis symmetry. At even higher temperatures all parameters undergo qualitative changes that reveal a pronounced coupling between magnetoelastic effects and magnetic anisotropy. This appears to provide experimental evidence for hole-mediated magnetostriction in dilute magnetic semiconductors. Strain may play a key role in the carrier modulation process, impacting both magnetostriction and magnetic anisotropy, and also giving rise to substantial inverse magnetoelastic anisotropy fields. This raises the prospect for engineering DMS devices that rely on externally applied stress to achieve novel control and tunability of magnetic behavior.

This work was supported by DARPA under grant DSO/SPINS-MDA 972-01-1-0024. KDG acknowledges support as research assistant of the Research Fund Flanders (FWO).